# Wireless Intrusion Detection Systems (WIDS)

**Ibrahim Al shourbaji[1] , Rafat Al Amer[2]**

[1]Lecturer, Computer Networks Department  
Jazan University , Jazan, Saudi Arabia

[2]Lecturer, Information Systems Department  
Jazan University , Jazan ,Saudi Arabia

**Abstract** – The rapid proliferation of wireless networks and mobile computing applications has changed the landscape of network security,the wireless networks have changed the way business, organizations work and offered a new range of possibilities and flexibilities; It is clear that wireless solutions are transforming the way we work and live. Employees are able to keep in touch with their e-mail, calendar and employer from mobile devices, but on the other hands they introduced a new security threats appeared. While an attacker needs physical access to a wired network in order to gain access to the network and to accomplish his goals, a wireless network allows anyone within its range to passively monitor the traffic or even start an attack ,one of the countermeasures that can be used is the intrusion detection systems in order to allow us to know both the threats affecting our wireless network and our system vulnerabilities in order to prevent those attackers, the IDS its main purpose is to manage the system and its operations and Its duty depends on .

## 1. Introduction

Within this paper will briefly focus on WLAN network, the security threats for wireless networking, specifically focusing on intrusion detection systems Through wireless IDS.

The Wireless Local Area Networks, or WLANs, are defined by the IEEE 802.11 families of standards. An 802.11 WLAN consist of stations (laptops, PDAs, mobile phones etc) and access points (or APs), which logically connect the stations with a distribution system (DS), typically the organization's wired infrastructure. A WLAN can run in ad-hoc mode, without the use of APs, and involving a direct communication between stations and in infrastructure mode, in which case the station connects to a DS via the access point. The identification of stations and APs is made by the use of 48-bit MAC addresses. Protocols (like Wi-Fi Protected Access) were introduced, which offer a better protection, but still suffer from different security issues.

The adoption of the WLANs in organizations introduced new specific threats for them, and as we will see in this paper some of these issues can be detected by using wireless intrusion prevention systems.

Rogue access points - represent unauthorized access points and can be either internal or external. The internal rogue AP is connected to the wired network by an unauthorized user (such as a regular employee), outside the control of the IT personnel. It can behave like a gateway for an attacker who can gain access to the network without the need to be physically inside the organization's perimeter. Therefore the detection and the removal of such rogue access points must be considered a critical aspect. It can be noted that this threat can affect also organizations which do not use WLAN networks in their activity. The external rogue access point is not connected to the wired organization's intranet, but emulates a legitimate access point of the network. For example, the attacker can set the rogue access point's SSID to the same SSID like the legitimate AP, and then increase significantly the signal of the rogue AP .

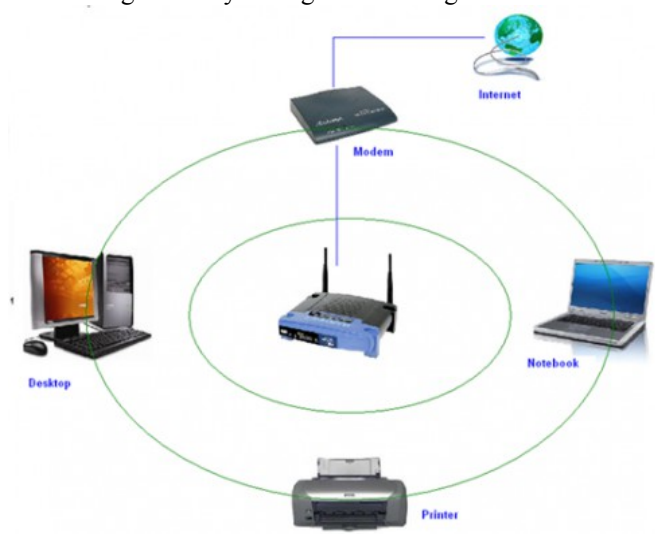

**Figure -1 wireless network**

By definition, wireless frequencies are designed to be heard by anyone with a wireless receiver – anyone can tune into a wireless network in the same way that they can tune into a radio station. It's simplicity makes wireless networks such a potential threat.

**2-Some of the most common threats against wireless network:**

**2.1-MAC address spoofing**. An access point can be configured so that it keeps a list of the legitimate client stations by MAC address. The attacker has the option of compromising such a client, or by spoofing with a legitimate MAC address. The MAC addresses are uniquely assigned at the time of manufacture, but usually this value can be set to arbitrary chosen values using an appropriate software tool

**2.2- Denial-of-Service**. A denial of service (DoS) attack occurs when a system cannot provide services to authorized clients due to resource exhaustion by unauthorized clients. This can be done by *jamming* (generate random signal on the specified frequencies), *flooding with associations* (the association table maintained by the AP has a maximum value – when this table overflows, the AP cannot accept a new further client association requests), *forged disassociation* (the attacker sends spoofed disassociation frames with the source MAC address of the AP – in this case the station is still authenticated but has to send a Reassociation request to the AP; to prevent reassociation, the attacker can continue to send disassociation frames for a specific period), *forged de authentication* (an attack similar to *forged disassociation*, but which uses de authentication frames).

*2.3- man-in-the-middle attacks* : are sometimes known as fire brigade attacks. The term derives from the bucket brigade method of putting out a fire by handing buckets of water from one person to another between a water source and the fire.

**2.4- Flooding** Attempts to flood the AP with Associations

**3-Wireless Intrusion Detection**

**In order to protect our network we need to ensure that we know**
-where all access points reside on our network
-what actions to take to close down any unauthorized access points or users
 -what wireless users are connected to our network
  - what unencrypted data is being accessed and exchanged by those users

it becomes obvious that for any organization using WLANs the monitoring of the air space should be an important measure in assuring a proper level of network security ,also we must monitor our air space using a Wireless Intrusion Detection systems

**4-What is Intrusion Detection Systems??**
First thing to clarify here is that an IDS is not a firewall!

Firewalls are designed to be outward looking and to limit access between networks in order to prevent an intrusion happening. But IDS watch the wireless and wired network from the inside and report or alarm depending on how they evaluate the network traffic they see and also it can identify and alert to the presence of unauthorized MAC addresses on the networks. This can be an invaluable aid in tracking down hackers.

An Intrusion Detection System (IDS) is a software or hardware tool used to detect
unauthorized access of a computer system or network .These systems monitor traffic on your network looking for and logging threats and alerting personnel to respond.

An IDS usually performs this task in one of two ways, with either signature-based or anomaly based detection.

Almost every IDS today is at least in part signature-based. Attacks and their tools
usually have a unique signature that can be detected and/or found. This means that known attacks can be detected by looking for these signatures. The downside to these is that they are easy to fool and can only detect attacks for which it has a signature.

**5-Why Use a Wireless Intrusion Detection System??**

The traditional wired IDS system does very little for the wireless world. The problem with wireless is that in addition to attacks that may be performed on a wired network, the medium itself has to be protected. To do this there are many measures which can be taken, however there are even more tools designed to break them.
Due to the nature of wireless LANs (WLAN), it can be difficult to control the areas of access. Often the range of a wireless network reaches outside the physical boundaries of an organization. With such a problem with wireless security, developing and implementing WIDS systems is definitely a step in the right direction.
If you have wireless and are concerned about attacks and intruders, a WIDS may be a great idea .A large number of possible attacks can be detected by a WIDS.
 The following will list major attacks and events that can be detected with the help of a WIDS. Rogue devices, such as an employee plugging in an unauthorized wireless router, incorrect configurations, connectivity problems, jamming, man-in-the middle  attacks, wardrivers, scanning with programs like Netstumbler or Kismet, RF interference, MAC spoofing, DoS attacks, attempts of brute force to get pass 802.1x, strong RFI, or use of traffic injection tools. Different WIDS devices and software have different capabilities in what can be detected. Make sure the WIDS you chose will fit your company's profile .There are currently only a handful of vendors who offer a wireless IDS solution - but the products are effective and have an extensive feature set. Popular wireless IDS solutions include Airdefense, RogueWatch and Airdefense Guard, and Internet Security Systems Realsecure Server sensor and wireless scanner products. A homegrown wireless IDS can be developed with the use of the Linux operating system, for example, and some freely available
software. Open source solutions include Snort-Wireless and WIDZ, among others.

**6-Some Wireless Encryption Standards**

Wireless encryption uses Wired Equivalent Privacy (WEP) standard to provide a secure communications, WEP uses a symmetric encryption scheme where a shared key is used for both encryption and decryption. It was, however, quickly breached and anyone intercepting and monitoring the wireless traffic could easily break the encryption using a brute force attack using tools such as WEP Crack .

The next wireless security standard introduced to try to bolster WEP was Wi-Fi
Protected Access or WPA. As an improved  a approach to key encryption by mixing the key

CISCO System's Lightweight Extensible Authentication Protocol (LEAP) is the
last encryption standard ,uses the authentication data to pass between the AP
and a Remote Authentication Dial in Service (RADIUS). This mutual
Authentication helps mitigate eavesdropping and man-in-the-middle attacks.

**7-Choosing a Wireless Intrusion Detection System
We need to decide which WIDS to implement and how , architecture of a wireless IDS , and general overview of Commercial WIDS systems vs. Open Source WIDS systems.**

A wireless IDS can be deployed in one of two ways centralized or decentralized.
In a decentralized environment each WIDS operates independently, logging, and alerting on its own. In addition this also means each WIDS has to be administered independently

In a large network this can quickly become overwhelming and inefficient, and therefore is not recommend for networks with more than one or two access points. The idea behind a centralized WIDS is that sensors are deployed that relate information back to one central point. This one point would send alerts and log events as well as serve as a single point of administration for all sensors. Another advantage to a centralized approach is that sensors can collaborate with one another in order to detect a wider range of events with more accuracy.

In this approach there are also three main ways in which sensors can be deployed

- **by using existing access points (AP)**
  Some access points on the market are able to simultaneously function as an AP and WIDS sensor.

This option has the potential to be less expensive than the others however there is a downside. Using the AP for both functions will reduce the performance, potentially creating a "bottle neck" on the network.

- **to deploy "dumb" sensors**
  These devices simply relay all information to the central server and rely
  on the server to detect all events. While inexpensive, all information is sent back to a central point causing an impact in the performance of the wired network and creating a single point of failure at the server

- **to use of intelligent sensors**
  These devices actively monitor and analyze wireless traffic, identify attack patterns and rouge devices as well as look for deviations from the norm. They then report these events back to the central server and allow an administrator to invoke countermeasures.

WIDS consists of :agent ,sensor .server ,consol and management ,reporting tools

Wireless IDS systems are available either as a complete hardware/software
solution or as a software only solution. An example of one such commercial hardware device is AirDefense Guard (www.airdefense.net).

- Commercial systems are expensive and , can lack some configuration abilities, but they tend to be more of an out-of-the-box deployment, with less knowledge and work needed to get started. Commercial systems normally provide more technical support along with a more user friendly interface for configuration, monitoring, and reporting.

One of the biggest disadvantages to many commercial sensors is the inability to change the antenna.

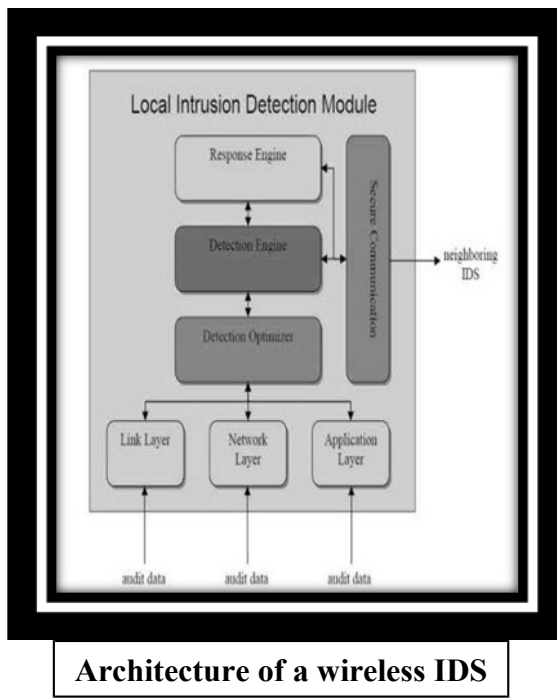

**Architecture of a wireless IDS**

- Open Source solutions provide many options and worlds of flexibility. These systems tend to work logically the same as the commercial solutions. They give you freedom to install on the hardware of choice. You also have more flexibility in the configuration of the software itself. Open source options are free with the exception of hardware and allow unlimited possibilities for installation, from modifying program functionality to custom hardware

However it often takes more time and effort along with a deeper knowledge to correctly install and configure open-source systems. Some examples include: AirIDS, Kismet, and SNORT-Wireless.

It is up to you to decide which will be the best solution for your network and what is the risk assessment it will cause in case an attacks happened. what will be affecting and what the proper should be taken before ,during and after the attack happened.

We can also implement the following steps for effective response:
Preparation, Identification, Initial response, Formulation of response strategy, Investigation of the incident, Reporting, Resolution and risk documentation

There is never one solution that works for everything.

# 8-IDS Implementation and Maintenance

## 8.1-Implementation

Once a wireless IDS product has been selected, the administrators need to design architecture, perform IDS component testing, secure the IDS components, and then deploy them. Implementing a wireless IDS can necessitate brief wireless network outages if existing APs or wireless switches need to be upgraded or have IDS software installed. Generally, the deployment of sensors causes no network outages.

## 8.2-Operation and Maintenance

Wireless IDS consoles offer similar management, monitoring, analysis, and reporting capabilities. One significant difference is that wireless IDPS consoles can display the physical location of threats. A minor difference is that because wireless IDS sensors detect a relatively small variety of events, compared to other types of IDSs, they tend to have signature updates less frequently.

### Enforcing a wireless policy

Creating and enforcing a wireless policy is the most important aspect of wireless security. Without policy anything goes.

A wireless IDS not only detects attackers, it can also help to enforce policy. WLANs have a number of security-related issues, but many of the security weaknesses are fixable. With a strong wireless policy and proper enforcement, a wireless network can be as secure as the wired equivalent - and a wireless IDS can help with the enforcement of such a policy.
Suppose policy states that all wireless communications must be encrypted. A wireless IDS can continually monitor the 802.11 communications and if a WAP or other 802.11 device is detected communicating without encryption, the IDS will detect and notify on the activity. If the wireless IDS is pre-configured with all the authorized WAPs and an unknown (rogue) WAP is introduced to the area, the IDS will promptly identify it. Features such as rogue WAP detection, and policy enforcement in general, go a long way to increase the security of the WLAN. The additional assistance a wireless IDS provides with respect to policy enforcement can also maximize human resource allocation. This is because the IDS can automate some of the functions that humans would ordinarily be required to manually accomplish, such as monitoring for rogue WAPs

### Limitations

Although a Wireless IDS can do a lot of things, it has its limitations. For example, it
cannot detect a passive sniffer – and usually an attacker can first collect data traffic before launching an attack. This period of passive sniffing is quite dangerous, but there is nothing to do in this direction. The only countermeasure is to use the proper protection through encryption..

Another limitation to NIDS manifests as bandwidth rates increase in a network. Especially when the amount of traffic also increases, it becomes a challenge for NIDS to be able to keep up with the rate of traffic and analyze data quickly and sufficiently.

Many networks are large and can even contain a heterogeneous collection of thousands of devices.  Sub- components in a large network may communicate using different technologies and protocols. One challenge for IDS devices deployed over a large network is for IDS components to be able to communicate across sub-networks, sometimes through firewalls and gateways..

**Summary**

The nature of a wireless network is to provide easy access to end users, but this ease of access creates a more open attack surface.

WLAN consist of stations (laptops, PDAs, mobile phones etc) and access points (or APs), which logically connect the stations with a distribution system (DS),

DoS , man- in- the middle attacks ,MAC address spoofing are the most common threats against wireless networks

Wireless has and is opening many new possibilities for expanding networks. Its
potential is amazing. As with most new technologies, wireless has several vulnerabilities.

WEP, WPA, Lightweight Extensible Authentication Protocol (LEAP) are some Wireless Encryption Standards

The importance of using an IDS and its ability to reduce the risk that may cause by attackers may gain access to your system, also it will draw picture for what is going on your wireless network

Luckily new developments like the Wireless IDS have come about that address many of these. Wireless IDS solutions are available from both the open-source and commercial markets and both have their pros and cons , It is up to you to decide which will be the best solution for your network. In any network with or without wireless never forget the creation and enforcement of policy.

 We need to decide which WIDS to implement, the administrators need to design architecture, perform IDS component testing, secure the IDS components, and then deploy them.

**CONCLUSION**

We have observed that any secure network will have vulnerability that an adversary could exploit by the attackers, this especially true for wireless networking.

 Intrusion detection can compliment to intrusion prevision techniques (such as encryption, authentication, secure routing ....)to improve the network security however the new techniques must  be developed to make intrusion detection work better for the wireless environment .

One main disadvantage that traditional signature-based intrusion detection systems have is that they cannot detect intrusions that are newly formed and depend largely on old known attacks. We have introduced an Intrusion detection to detect new attacks that are spawning daily in the hands of malicious hackers

In any network with or without wireless never forget the creation and enforcement of policy. There is never one solution that works for everything as a perfect solution for the problems in wireless networks but is, we believe, a little better over the rest , and IDS is only one part of a greater security solution for wireless network .

Finally The IDS must be able to recognize the different data formats and different protocols for communication.